\documentclass[pdflatex,sn-mathphys-num]{sn-jnl}

\usepackage{graphicx}
\usepackage{multirow}
\usepackage{amsmath,amssymb,amsfonts}
\usepackage{amsthm}
\usepackage{mathrsfs}
\usepackage[title]{appendix}
\usepackage{xcolor}
\usepackage{textcomp}
\usepackage{manyfoot}
\usepackage{booktabs}
\usepackage{algorithm}
\usepackage{algorithmicx}
\usepackage{algpseudocode}
\usepackage{listings}
\usepackage{minted}
\usepackage{dblfloatfix}
\usepackage{ulem}
\usepackage{caption}
\usepackage{subcaption}
\usepackage{dramatist}
\usepackage{xspace}
\usepackage{pifont} 
\usepackage{tcolorbox}
\usepackage{xltabular}
\usepackage{longtable}
\usepackage{lineno}
\usepackage{adjustbox}
\usepackage[symbol]{footmisc}
\usepackage{setspace}
\usepackage{cleveref}
\usepackage{dsfont}
\usepackage{array}
\usepackage{tabularx}
\usepackage{lipsum}  
\usepackage{multicol}
\usepackage{titlesec}
\usepackage{lineno}
\usepackage{hyperref}

\titleformat{\paragraph}[runin]{\bfseries\normalsize}{}{0pt}{}[]
\titlespacing*{\paragraph}{0pt}{0.5\baselineskip}{0.75em}
\usepackage{booktabs}   
\usepackage{makecell}   
\usepackage{pifont}     
\usepackage{siunitx}        
\usepackage{fontawesome5}   

\theoremstyle{thmstyleone}%
\theoremstyle{thmstyletwo}%

\theoremstyle{thmstylethree}%

\begin{document}


\author[1]{\fnm{Shimin} \sur{Di}}\email{shimin.di@seu.edu.cn}
\equalcont{These authors contributed equally to this work.}

\author[2]{\fnm{Xujie} \sur{Yuan}}\email{yuanxj8@mail2.sysu.edu.cn}
\equalcont{These authors contributed equally to this work.}

\author[1]{\fnm{Hanghui} \sur{Guo}}\email{ghh1125@zjnu.edu.cn}
\equalcont{These authors contributed equally to this work.}

\author[2]{\fnm{Chaoqian} \sur{Ouyang}}\email{ouychq@mail2.sysu.edu.cn}

\author[1]{\fnm{Yongxu} \sur{Liu}}\email{yongxuliu@seu.edu.cn}

\author[3]{\fnm{Ling} \sur{Yue}}\email{yuel2@rpi.edu}

\author[4]{\fnm{Zhangze} \sur{Chen}}\email{zjnuczz@zjnu.edu.cn}

\author*[2]{\fnm{Libin} \sur{Zheng}}\email{zhenglb6@mail.sysu.edu.cn}

\author[4]{\fnm{Jia} \sur{Zhu}}\email{jiazhu@zjnu.edu.cn}

\author[3]{\fnm{Shaowu} \sur{Pan}}\email{pans2@rpi.edu}

\author[2]{\fnm{Jian} \sur{Yin}}\email{issjyin@mail.sysu.edu.cn}

\author[1]{\fnm{Yong} \sur{Rui}}\email{yrui@acm.org}

\author*[1]{\fnm{Min-Ling} \sur{Zhang}}\email{zhangml@seu.edu.cn}






\affil[1]{\orgname{Southeast University}, \city{Nanjing}, \country{China}
}

\affil[2]{\orgname{Sun Yat-sen University}, \city{Zhuhai}, \country{China}
}

\affil[3]{\orgname{Rensselaer Polytechnic Institute}, \city{Troy}, \country{USA}
}

\affil[4]{\orgname{Zhejiang Normal University}, \city{Jinhua}, \country{China}
}


\newcommand{\zdaxie}[1]{\textcolor[rgb]{0.545,0,0.071}{#1}}
\newcommand{\yd}[1]{{{\textcolor{cyan}{[Yadi: #1]}}}}
\newcommand{\sw}[1]{{{\textcolor{red}{[Shaowu: #1]}}}}
\newcommand{\fa}{Foam-Agent\xspace}
\newcommand{\fb}{CFDLLMBench\xspace}


\title{\texorpdfstring{\raisebox{-0.7em}{\includegraphics[height=2.5em]{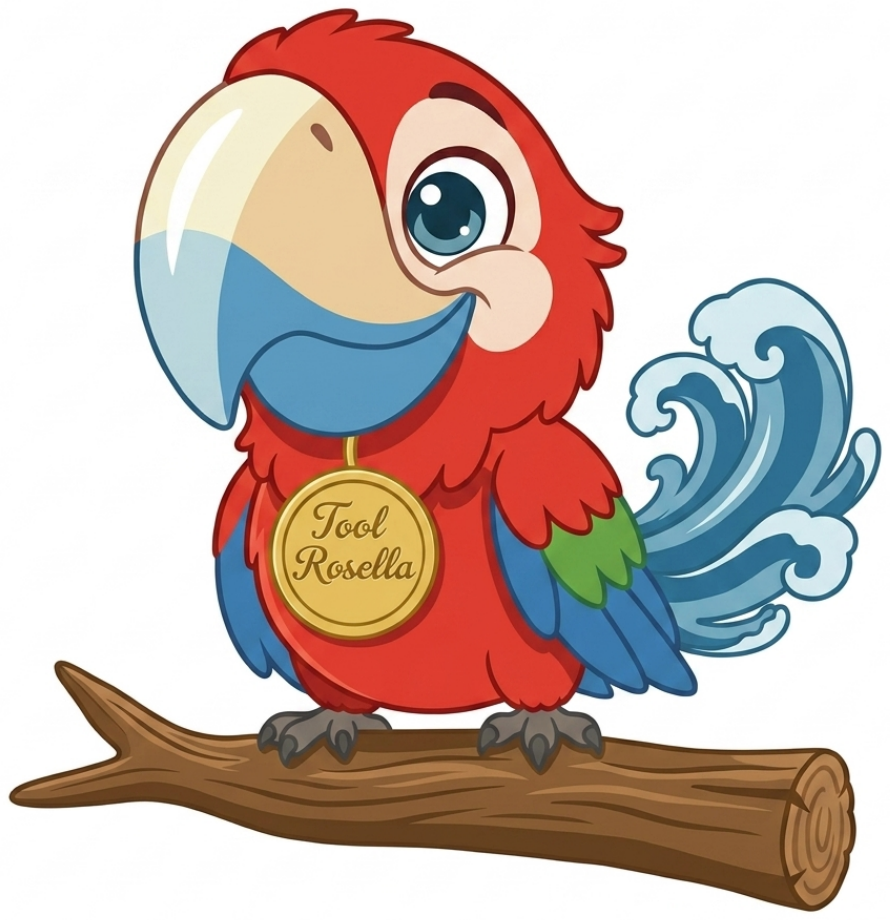}}\ }{}ToolRosella: Translating Code Repositories into Standardized Tools for Scientific Agents}

\abstract{
Large Language Model (LLM)-based agent systems are increasingly used for scientific tasks, yet their practical capability remains constrained by the narrow scope of manually curated tools they can invoke. Much scientific computational functionality already exists in open-source code repositories, but these resources remain difficult to standardize, operationalize, and invoke reliably for agent use. Here we present ToolRosella, a framework that automatically transforms heterogeneous scientific code repositories into standardized, agent-invocable tools. ToolRosella combines repository analysis, tool interface construction, execution testing, and iterative repair to address the problem of repository-to-tool standardization. 
Across 122 GitHub repositories spanning 35 subdisciplines in six domains, ToolRosella reaches a 61.5\% repository conversion success rate after iterative repair, with a 4.4× speedup over human engineers. The resulting 1,580 callable tools support a downstream task success rate of 84.0\% and improve performance when integrated into other agent frameworks, particularly on tasks whose required tools are absent from fixed, curated inventories.
}

\keywords{AI for Science, Large Language Model, Agentic AI, Tool Standardization}

\maketitle

\section{Introduction}

Artificial Intelligence for Science (AI4Science)~\citep{wang2023scientific,xu2021artificial} is shifting from a predominantly model-centric paradigm toward a more agentic one~\cite{swanson2025virtual}. Earlier AI4Science systems are often designed to address individual tasks within scientific workflows, such as prediction~\citep{jumper2021highly,bi2023accurate} or simulation~\citep{hayes2025simulating,hatfield2021data}. More recently, large language models (LLMs)~\citep{lin2023evolutionary,qu2024promoting} have made it increasingly feasible to build agentic systems for more complex scientific tasks~\citep{xin2025towards,lu2026towards}.
These tasks, however, ultimately reduce to concrete computational operations, including data processing, structure analysis, simulation, optimization, and experimental design, which are implemented and exposed through code repositories, software packages, and APIs~\cite{tee2022framework}. Recent scientific LLM-based agents, including ChemCrow~\citep{MBran2024}, SciSciGPT~\cite{shao2025sciscigpt}, and Coscientist~\citep{Boiko2023}, illustrate that what makes such systems practically useful is their ability to invoke, orchestrate, and execute external tools. Tool invocation is therefore a key operational mechanism through which agents obtain executable scientific capability~\cite{vriza2026operating}.

\begin{figure}[!t]
\centering
\includegraphics[width=\linewidth]{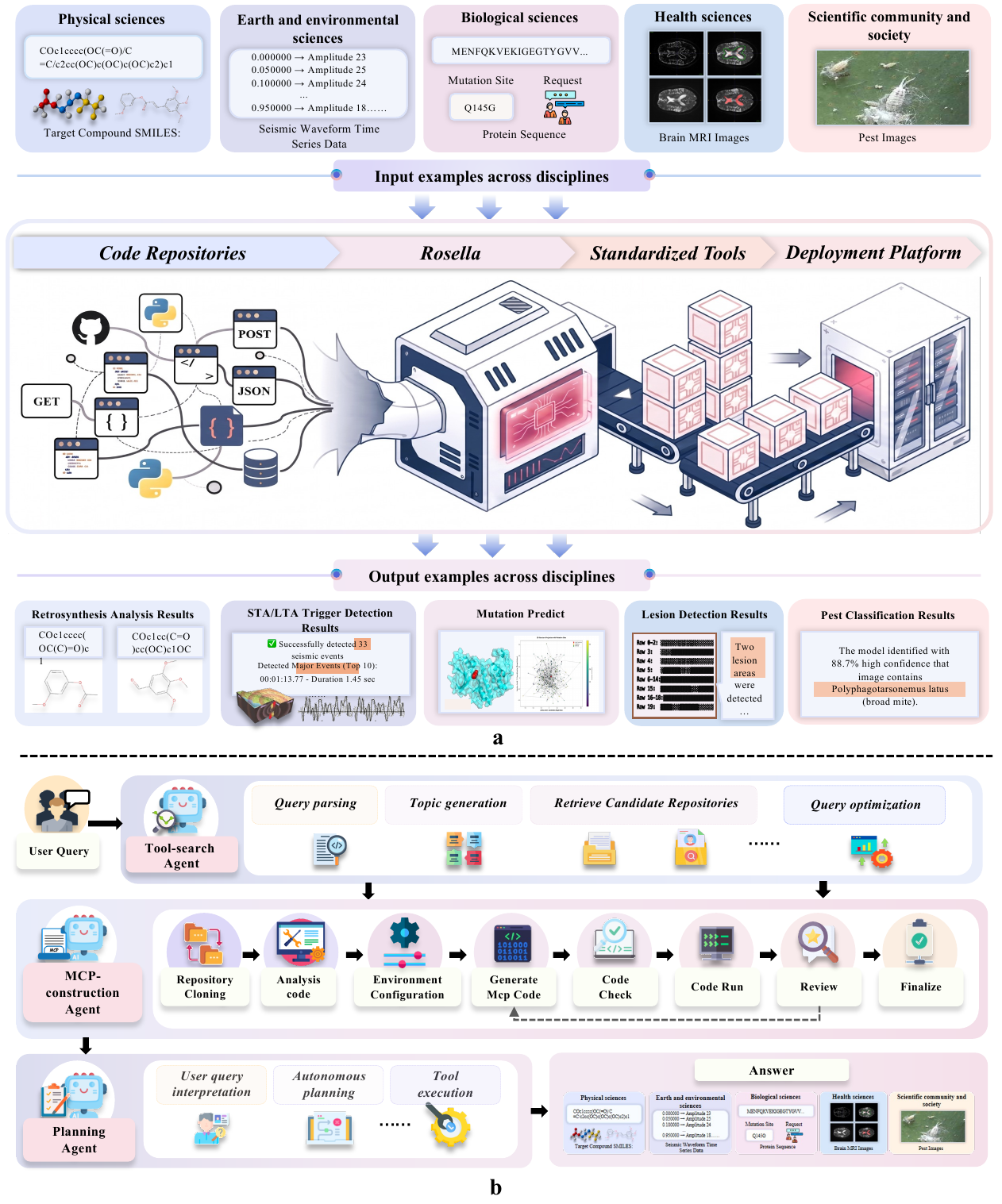}  
\caption{Overview of ToolRosella:
\textbf{a}, ToolRosella's strength lies in how to automatically transform heterogeneous scientific resources into callable tools, bridging existing repositories and agent execution. Unlike fixed, manually curated tool libraries, it focuses on repository-to-tool standardization for reusable tool construction.
\textbf{b}, The pipeline for using ToolRosella to complete scientific tasks.
}
\label{fig:ToolRosella_Framework}
\end{figure}

Existing approaches to scientific tool use broadly follow two routes, each with a different limitation. One route emphasizes reliability by operating over a curated and predefined tool inventory, such as the 500+ tools used in SciToolAgent~\citep{Ding2025}, as well as smaller but carefully designed toolsets in systems like SciSciGPT~\cite{shao2025sciscigpt}, using 11 tools, and CRISPR-GPT~\cite{qu2026crispr}, using 4 search-based tools. These systems perform well because useful tools have already been manually selected, standardized, and integrated in advance, but this also constrains the agent to a fixed and precompiled capability space. Another route attempts to leverage larger external repository and tool ecosystems, as in systems such as RepoMaster~\cite{wang2025repomaster} and OpenAgents~\cite{lyu-etal-2025-enhancing}. However, resources in these ecosystems are highly heterogeneous. They are developed independently, with diverse interfaces, environments, and calling conventions~\cite{ferber2025development,yang2026tool} that are not naturally aligned for direct agent use~\cite{yang2024swe}. As a result, broader tool coverage often comes at the cost of lower invocation reliability. The missing piece between these two routes is a standardization layer that can convert heterogeneous scientific repositories into callable tools for agent execution.

In practice, much of the computational functionality needed for scientific tasks is already implemented in existing open-source repositories~\cite{woelfle2011open,van2024encore}. Yet these resources are rarely available to agents in a directly usable form~\cite{di2025stop}. They are difficult to operationalize because of inconsistent interfaces, fragile dependency environments, incomplete execution specifications, and highly variable implementation styles. As a result, a large portion of scientific tools remains effectively locked away from LLM-based agents, even when the underlying code already contains the computational capability needed for a given task.  The core challenge, therefore, is not merely locating relevant scientific code but transforming heterogeneous and previously non-callable programs into tools that agents can use reliably~\cite{wang2026ai}. 

We address this challenge with ToolRosella, a framework that automatically transforms heterogeneous scientific code repositories into standardized, agent-invocable tools (Fig.~\ref{fig:ToolRosella_Framework}a), implemented as a hierarchical multi-agent pipeline (Fig.~\ref{fig:ToolRosella_Framework}b). Rather than operating over a fixed library of manually prepared tools, ToolRosella focuses on the upstream problem of repository-to-tool standardization. In our implementation, this standardization is realized through \href{https://blog.modelcontextprotocol.io/posts/2025-11-25-first-mcp-anniversary/}{Model Context Protocol (MCP)}-compatible service construction. ToolRosella analyzes relevant repositories, constructs standardized tool interfaces, validates executability through testing, and iteratively repairs failures when they arise. In this way, it bridges heterogeneous scientific software and downstream agent execution, making existing computational functionality more directly reusable by agent systems. 
The main idea of ToolRosella is to provide a new infrastructure layer that makes previously inaccessible scientific tools available to agents as executable capabilities.

We make three main contributions. First, we present an end-to-end framework for repository-to-tool standardization that automatically transforms scientific code repositories into callable tools for LLM-based agents. Second, on 122 GitHub repositories spanning 35 subdisciplines and six domains, ToolRosella standardizes 1,580 callable tools and reaches a 61.5\% repository conversion success rate after iterative repair while substantially reducing average standardization time relative to human engineers. Third, these standardized tools form a reusable capability layer for downstream agents. With the full tool pool constructed from all 122 repositories, ToolRosella achieves an average task success rate of 84.0\% across domains, and the same tool layer can be integrated into other agent frameworks, such as \href{https://openclaw.ai/}{OpenClaw}, to improve task completion, especially when required tools are absent from fixed, curated inventories.


\section{Results}
\subsection{Overview of ToolRosella}
\label{subsec:overview-of-toolrosetta}

\begin{figure}[b]
\centering
\includegraphics[width=\linewidth]{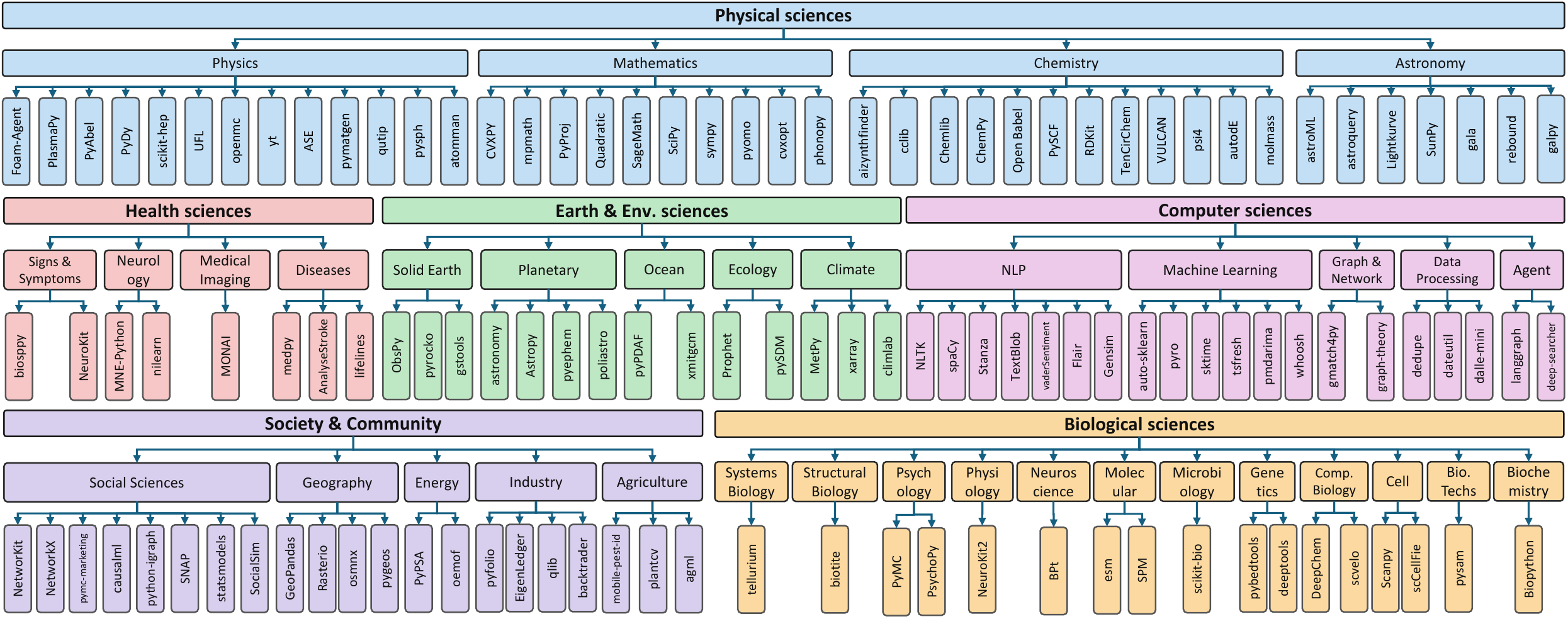}  
\caption{Overview of the ToolRosella ecosystem. Each node represents a GitHub repository automatically converted into standardized tool services, organized into five major scientific areas (Physical Sciences, Earth \& Environmental Sciences, Biological Sciences, Health Sciences, and Scientific Community \& Society) and Computer Science. Node groupings reflect sub-domain categorization within each area.}
\label{fig:ecosystem}
\end{figure}

As illustrated in Figure~\ref{fig:ToolRosella_Framework} (b), ToolRosella implements a hierarchical multi-agent architecture comprising three core components. 
A \textit{Tool-search Agent} retrieves and evaluates candidate repositories through LLM-driven semantic parsing of user queries, GitHub API retrieval ranked by repository popularity, and assessment of each candidate's code completeness and functional relevance. 
An \textit{MCP-construction Agent} transforms qualifying repositories into standardized services through an automated pipeline that combines code analysis, environment setup, service generation, and iterative validation and repair.
Finally, a \textit{Planning Agent} operates after tool construction and registration, interpreting user queries to autonomously select, schedule, and invoke the most suitable tools from the constructed toolset to generate the final answer.
As shown in Figure~\ref{fig:ecosystem}, through multi-agent collaboration, ToolRosella has successfully translated 1,580 tools from 122 GitHub repositories covering 5 major scientific areas (Physical Sciences, Earth \& Environmental Sciences, Biological Sciences, Health Sciences, Scientific Community \& Society)
and Computer Science.


\subsection{How well does automated tool standardization work?}
\label{subsec:standardization-performance}
Standardizing open-source repositories into callable tool services has traditionally required substantial manual effort from trained engineers. ToolRosella aims to automate this process. To evaluate its performance, we benchmark repository-level conversion on 122 GitHub repositories spanning 35 subdisciplines and six domains. We compare ToolRosella's initial conversion round (hereafter \textit{``first-pass"}, i.e.\ before the Review-Revise-Fix repair loop) against human coding engineers (assisted by GPT-5.3-Codex) and an LLM-only baseline (GPT-5.3-Codex) that directly generates the complete tool service stack in a single pass. Success is defined as exposing at least three validated tool endpoints that an agent can correctly invoke with valid outputs.

\begin{figure}[t]
\centering
\includegraphics[width=\linewidth]{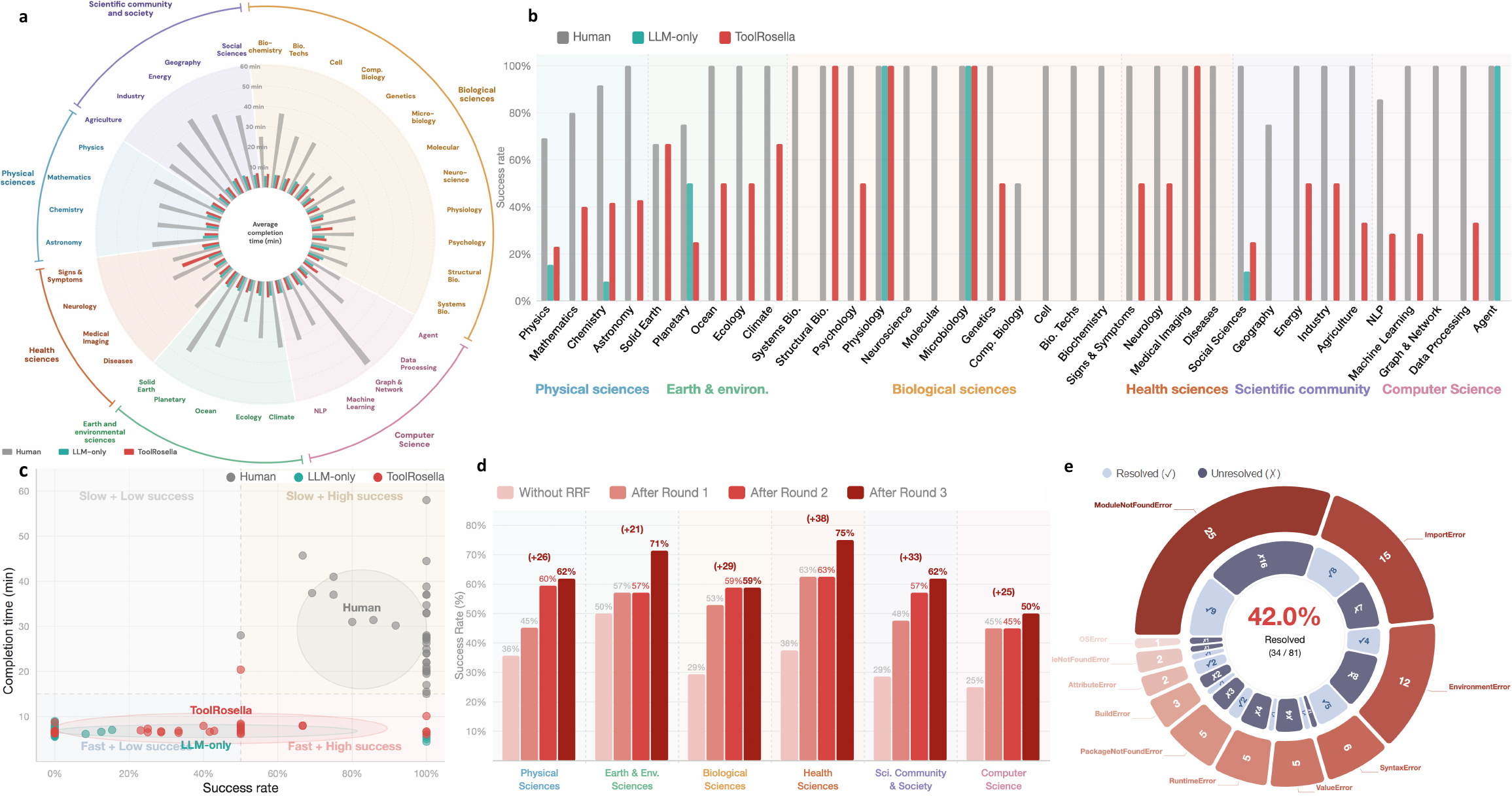}  
\caption{Automated tool conversion performance evaluation. \textbf{a}, Radial bar chart of average completion time per subdiscipline (in minutes) for ToolRosella, human experts, and the LLM-only baseline. \textbf{b}, Repository conversion success rates across 35 subdisciplines within six domains, comparing all three methods. \textbf{c}, Joint visualization of success rate versus completion time, where each point represents one subdiscipline. \textbf{d}, Repair-focused ablation showing cumulative improvement from successive rounds of the Review-Revise-Fix (RRF) mechanism across six domains. \textbf{e}, Error types and their resolution statistics after three RRF rounds.}
\label{fig:fig_results_2.2}
\end{figure}

\textbf{Overall conversion performance of efficiency and effectiveness.}
ToolRosella completes conversion in approximately 7.2\, min per repository on average, compared with 31.6\, min for Codex-assisted human engineers, a 77.2\% reduction in standardization time and a 4.4$\times$ speedup (Fig.~\ref{fig:fig_results_2.2}a,\,c). The LLM-only baseline is marginally faster (approximately 7\,min), yet this speed advantage is eclipsed by its substantially lower reliability. As shown in Figure~\ref{fig:fig_results_2.2}(b), the LLM-only baseline achieves only 8.2\% (10/122), suggesting that single-pass generation alone is insufficient to handle end-to-end repository standardization. ToolRosella reaches 33.6\% (41/122) on the first pass and 61.5\% (75/122) after three rounds of iterative repair, compared with 90.2\% (110/122) for human engineers. These results indicate that iterative construction and repair substantially improve reliability over single-pass generation while maintaining a low average conversion time.

Performance varies across domains. In terms of efficiency, ToolRosella maintains relatively stable conversion time across domains (6--8\,min), whereas human engineers exhibit high variance, from 19.5\,min to 58.0\,min (Fig.~\ref{fig:fig_results_2.2}a), reflecting differences in repository complexity and dependency depth. In terms of success rate (Fig.~\ref{fig:fig_results_2.2}b), ToolRosella is strongest in Earth \& Environmental Sciences (50.0\%) and Health Sciences (37.5\%), followed by Physical Sciences (35.7\%), Biological Sciences (29.4\%), Scientific Community \& Society (28.6\%), and Computer Science (25.0\%). The LLM-only baseline exceeds 10\% in only two domains, and achieves 0\% in Health Sciences. This pattern suggests that ToolRosella’s gains over the LLM-only baseline are associated with its handling of environment construction, dependency resolution, and multi-step validation, rather than single-pass code generation alone.

\textbf{Iterative repair and Failure analysis.}
To recover first-pass failures, ToolRosella employs a multi-round Review-Revise-Fix (RRF) mechanism that diagnoses errors and applies targeted repairs. As shown in Figure~\ref{fig:fig_results_2.2}(d), the success rate rises from 33.6\% to 61.5\% after three rounds of repair, an absolute gain of 27.9 percentage points (pp). Most gains accrue in the first round (+15.6 pp), with diminishing returns thereafter. The largest single-domain improvements occur in Health Sciences (+37.5 pp) and Scientific Community \& Society (+33.3 pp), where complex dependency configurations and workflow-heavy repositories create low initial baselines that are nevertheless partially recoverable.
Of the 81 repositories that fail in the first pass, 34 are recovered after three RRF rounds while 47 remain unresolved (Fig.~\ref{fig:fig_results_2.2}e). Analyzing the deployment-stage exceptions that cause these failures reveals two broad groups. Dependency and environment issues dominate, accounting for 61/81 (75.3\%). \texttt{ModuleNotFoundError} alone constitutes the largest category (25/81, 30.9\%), followed by \texttt{ImportError} (15/81, 18.5\%), \texttt{EnvironmentError} (12/81, 14.8\%), and \texttt{PackageNotFoundError} (5/81, 6.2\%), with \texttt{BuildError} (3/81) and \texttt{OSError} (1/81) contributing smaller shares. The remaining 20/81 (24.7\%) are runtime and code-generation errors, including \texttt{SyntaxError} (6/81, 7.4\%), \texttt{ValueError} (5/81, 6.2\%), \texttt{RuntimeError} (5/81, 6.2\%), \texttt{AttributeError} (2/81, 2.5\%), and \texttt{FileNotFoundError} (2/81, 2.5\%). These results suggest that heterogeneous dependency environments remain a major bottleneck in automated repository-to-tool standardization.



\subsection{How useful are standardized tools for Agents?}
While Section~\ref{subsec:standardization-performance} demonstrates ToolRosella's capability to successfully standardize repositories into tools, the ultimate goal is to enable agents to leverage these standardized tools for solving real tasks. These two evaluations target different stages. Section~\ref{subsec:standardization-performance} measures the success rate of automated repository-to-tool conversion itself, whereas this section measures downstream task completion given a standardized tool layer. ToolRosella in the downstream evaluation uses the full tool pool constructed from all 122 repositories, combining the automatically converted with the remaining repositories completed by human engineers. We compare ToolRosella with five representative agent systems spanning distinct technical paradigms in tool-augmented agents. SciToolAgent~\cite{Ding2025} represents curated scientific tool use with manually prepared interfaces. ChemCrow~\citep{MBran2024} represents a domain-specialized agent with crafted prompting and chemistry-specific tools. RepoMaster~\citep{wang2025repomaster} and OpenAgents~\citep{lyu-etal-2025-enhancing} represent direct repository understanding and execution without an explicit standardization layer. OpenClaw represents a general-purpose autonomous agent framework with plugin-based tool orchestration. 
We report two metrics. \textit{Execution Success} measures whether a system can successfully produce an output, regardless of whether the final result is correct. \textit{Verified Success} applies a stricter criterion. A task is counted as successful only if the produced output is further judged by an independent LLM-based verifier (Claude Opus 4.6), with human review for ambiguous cases.

\begin{figure}[b]
\centering
\includegraphics[width=\linewidth]{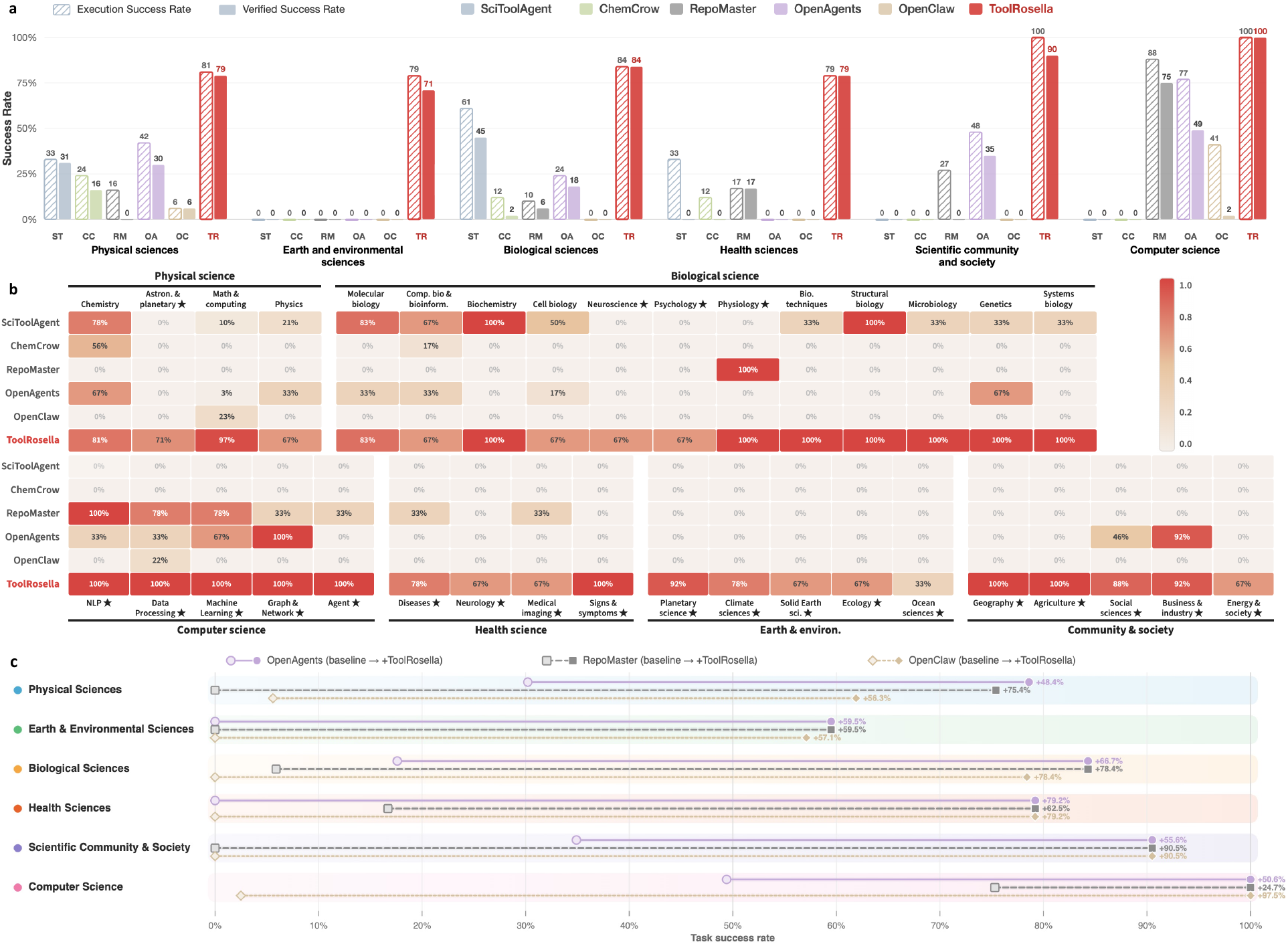}  
\caption{Downstream task evaluation results. \textbf{a}, Execution success rate and verified success rate of ToolRosella and five baseline systems (SciToolAgent, ChemCrow, RepoMaster, OpenAgents, and OpenClaw) across the six domains. \textbf{b}, Verified task completion success rates across 35 subdisciplines spanning the six categories. Stars denote 23 out-of-distribution (OOD) subdisciplines that are not covered by curated baseline tool sets. \textbf{c}, Verified performance gains when integrating ToolRosella-converted tools into three existing agent frameworks (OpenAgents, RepoMaster, and OpenClaw), with percentage annotations indicating improvement per domain.}
\label{fig:fig_results_2.3}
\end{figure}

\textbf{Task-Solving Performance Across Scientific Domains.}
Figure~\ref{fig:fig_results_2.3}(a) summarizes execution success and verified success across the six scientific domains. ToolRosella reaches 81\% execution success in Physical Sciences, 79\% in Earth \& Environmental Sciences, 84\% in Biological Sciences, 79\% in Health Sciences, and 100\% in both Scientific Community \& Society and Computer Science. Under the stricter verified criterion, ToolRosella maintains consistently strong performance across domains, achieves the highest verified success rate in all six domains, with an average verified success rate of 84.0\%. 

We further examine fine-grained verified success rate across 35 subdisciplines (Fig.~\ref{fig:fig_results_2.3}b). In this paper, we use ``out-of-distribution (OOD) subdisciplines'' as shorthand for subdisciplines that are not covered by the fixed, manually curated tool inventories of curated baselines. On these subdisciplines, ToolRosella achieves an average verified success rate of 82.5\%, compared with 0.0\% for both SciToolAgent and ChemCrow, 21.3\% for RepoMaster, 16.1\% for OpenAgents, and 1.0\% for OpenClaw. This gap underscores a limitation of fixed tool inventories and general-purpose agent frameworks. Even when baseline systems can reason about a task, they cannot execute the required computation to complete it if the relevant tools are unavailable.

\textbf{Benefits of Standardized Tools for Other Systems.}
To examine whether the gains are associated with the standardized tools themselves rather than only with ToolRosella's agent pipeline, we conduct a controlled augmentation experiment. We inject ToolRosella-converted tools into RepoMaster, OpenAgents, and OpenClaw while preserving each system's original architecture, prompting strategy, and reasoning pipeline. As shown in Figure~\ref{fig:fig_results_2.3}(c), all three systems improve substantially under this setting. RepoMaster increases from 16.3\% to 81.5\% (+65.2 pp), OpenAgents from 22.0\% to 82.0\% (+60.0 pp), and OpenClaw from 1.3\% to 77.9\% (+76.5 pp) in the average success rate. Improvements are observed across all six domains, with especially large gains in categories that were previously weak or uncovered by the original tool inventories. These results suggest that the value of ToolRosella lies not only in end-to-end task execution but also in providing standardized tool services that can be reused by different agent frameworks. Once constructed, these tools function as a transferable capability layer that broadens the operational tool space available to downstream agents.

\subsection{Case studies}

\begin{figure}[!t]
\centering
\includegraphics[width=\textwidth]{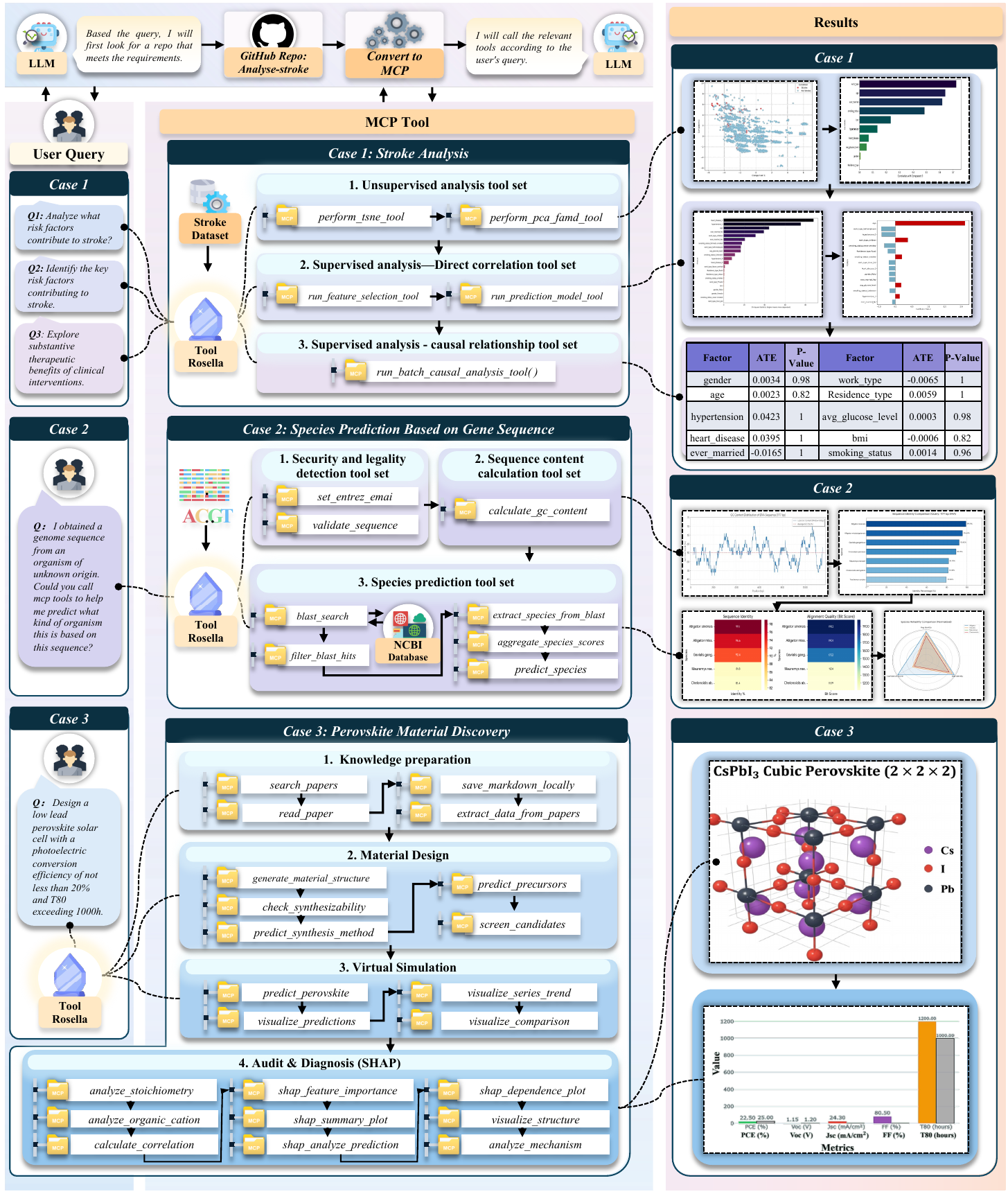}  
\caption{
Multi-domain case studies of ToolRosella for scientific workflow assistance. The framework converts user queries into executable MCP-compatible toolchains, enabling coordinated orchestration of task-specific analytical modules. Case 1 illustrates a stroke-analysis workflow integrating unsupervised, predictive, and causal inference tools, together with feature-importance summaries and treatment-effect estimates. Case 2 illustrates sequence-analysis workflows combining validation, compositional analysis, and database-driven species prediction. Case 3 illustrates a perovskite material design-and-evaluation workflow spanning literature mining, candidate generation, virtual screening, and mechanistic interpretation.}
\label{fig:case}
\end{figure}

To further validate the practical utility of ToolRosella as a scientific-assistance system in real-world research settings, we present three representative case studies spanning clinical data analysis, bioinformatics inference, and materials design (Fig.~\ref{fig:case}). Beyond evaluating the system’s ability to execute workflows, these cases highlight how ToolRosella assists users in constructing task-relevant tools, coordinating them into coherent pipelines, and supporting end-to-end scientific workflows in diverse domains.


\subsubsection{Case 1: Stroke Analysis}

Stroke analysis is an important task in clinical and biomedical research, with relevance to early diagnosis, risk assessment, and personalized treatment planning~\cite{kelly2024age, howard2025association}.
In this scenario, 
we apply ToolRosella to a clinical stroke dataset comprising structured patient features, including demographic variables and clinical measurements. Given a sequence of user queries, ToolRosella assists users by retrieving and standardizing task-relevant tools from open-source repositories, including data preprocessing modules, dimensionality reduction and visualization methods (for example, PCA and t-SNE), predictive modeling components (for example, logistic regression and tree-based models), and causal analysis libraries. These tools are then dynamically assembled into a unified workflow, reducing the need for manual integration.

The analysis proceeds in multiple stages guided by user queries. In response to the initial prompt, ToolRosella assists users by applying unsupervised methods to characterize latent data structure through dimensionality reduction and visualization. As subsequent prompts refine the analytical objective, the system updates the workflow and transitions to predictive modeling, where feature selection is used to quantify the contribution of candidate variables. Upon further queries seeking a deeper understanding of variable relationships, ToolRosella extends the workflow to causal analysis, estimating potential causal effects among key features. 
It identifies well-established stroke risk patterns, including higher risk in males than in females, increasing risk with age, and associations with hypertension and heart disease~\cite{kelly2024age,zhu2025global}. In addition, metabolic and lifestyle-related factors such as elevated blood glucose, obesity, and smoking emerge as important contributors within the resulting analysis~\cite{howard2025association}. Beyond these primary factors, the system also surfaces variables related to social and occupational context, such as marital status and employment type~\cite{sacco2024prevention}. These factors may influence stroke risk indirectly through psychosocial stress and lifestyle conditions rather than acting as direct biological causes. Some weaker signals (for example, urban residence) are also detected but show limited consistency, suggesting potential noise or spurious correlations in the data.


\subsubsection{Case 2: Species Prediction Based on Gene Sequence}

Species prediction is important in biological, agricultural, and ecological research, with applications in unknown-species identification, ecosystem diversity assessment, and the development and use of agricultural and microbial resources~\cite{xu2022fusarium, klughammer2023comparative}.
In this case, ToolRosella is applied to build a sequence-analysis pipeline from a user-provided gene sequence. Given an input sequence, the system assists users by retrieving and integrating task-relevant bioinformatics tools, including sequence validation modules, statistical feature-extraction utilities, sequence-alignment algorithms, and database-retrieval interfaces. These tools are dynamically assembled into a task-specific workflow with minimal manual configuration.

The analysis proceeds in multiple stages. ToolRosella first performs sequence validation and basic statistical analysis to extract compositional features such as nucleotide distribution and sequence length. It then conducts similarity-based retrieval against external biological databases, generating a set of candidate species on the basis of sequence-alignment scores. Building on the initial retrieval results, the system aggregates evidence from multiple sources and re-ranks candidate species using combined similarity and consistency signals. The final output consists of a ranked list of candidate species, with \textit{Alligator sinensis} consistently appearing as the top-ranked candidate by a clear margin over alternative predictions. The prediction is supported by high sequence-similarity scores and consistent alignment across multiple database sources. Lower-ranked candidates exhibit weaker alignment signals or partial matches, indicating lower confidence in those assignments.
The results are supported by consistency checks across independent retrieval tools and by ranking stability across repeated runs. Compared with baseline approaches relying on single-tool retrieval or static pipelines, ToolRosella integrates multiple tools and evidence sources, yielding more stable and consistent prediction results. 

\subsubsection{Case 3: Perovskite Material Design and Evaluation}

Perovskite solar cells have become a central focus of third-generation photovoltaics owing to their favorable optoelectronic properties and potential for low-cost manufacturing~\cite{li2025stabilizing, wu2025resilience}. Consequently, the design and evaluation of new perovskite materials has become an important research objective.
In this case, ToolRosella is applied to design and evaluation of low-lead, high-efficiency perovskite materials. Given a user query, the system assists researchers by retrieving and coordinating specialized tools for literature analysis, candidate generation and performance prediction. It first extracts structured prior knowledge from relevant literature, including compositional patterns, synthesis conditions and reported performance metrics. Guided by this knowledge, ToolRosella generates candidate material structures using integrated design tools and applies feasibility filtering to eliminate chemically implausible configurations. The remaining candidates are then evaluated through simulation, where key performance indicators such as predicted Power-Conversion Efficiency (PCE) are estimated and used to rank candidate materials. Based on intermediate results, the system further refines candidate prioritization by analyzing relationships between structural features and predicted performance. 

The final output consists of a ranked set of candidate materials, among which a Sn--Pb mixed perovskite composition is identified as the most promising solution, achieving a 40\%--50\% reduction in lead content while maintaining a tandem-compatible bandgap of 1.23 eV. Performance prediction indicates a PCE of 16\%--19\% (with theoretical potential exceeding 22\%) and an estimated stability of T80 $\approx$ 500--800 h under encapsulated conditions. Subsequent wet-lab experimental validation yields a PCE of 17\%, consistent with the predicted range. Although this result does not yet meet the original design targets (PCE $\geq$ 20\% and T80 $>$ 1000 h), achieving simultaneously high efficiency, reduced lead content and operational stability in Sn--Pb perovskite systems remains a challenging materials-design problem~\cite{yang2025understanding}. These results demonstrate that the constructed workflow can reliably identify experimentally viable candidates and produce consistent prediction-to-experiment alignment.
Compared with conventional approaches that require manual coordination of literature review, simulation tools and experimental design, ToolRosella enables an automated and integrated design-and-evaluation workflow that reduces iteration cost and human intervention. 

\section{Discussion}
\subsection{Contributions and limitations}

ToolRosella studies an upstream bottleneck in scientific agents. Many scientific code repositories contain useful computational functionality, but lack standardized interfaces for reliable agent invocation. Our results show that repository-to-tool standardization is feasible on a diverse benchmark of 122 GitHub repositories spanning 35 subdisciplines and six domains. ToolRosella achieves a 61.5\% repository conversion success rate after iterative repair, while substantially reducing average standardization time relative to human engineers. The downstream experiments show that these standardized tools are useful beyond the conversion setting itself. Using the full tool pool constructed from all 122 repositories, ToolRosella achieves strong verified task performance across six domains, and the augmentation experiments further suggest that the resulting tool services can be reused by different agent frameworks. Taken together, these results support the view that automated repository-to-tool standardization can serve as a practical capability layer for scientific agents, especially when required computational functions are not already available in fixed, curated tool inventories.


The current system also has limitations. A gap remains between automated conversion and human expert completion. Our failure analysis shows that iterative self-correction cannot fully resolve two major sources of conversion failure. The dominant barrier is environment reconstruction, including broken remote dependencies, platform-specific configurations, and implicit system-state assumptions, which often resist code-level repair because they require reasoning beyond the repository itself. The remaining failures arise from code structure heterogeneity, especially in repositories designed for interactive exploration, such as Graphical User Interface (GUI) applications or notebook-based workflows, that lack the programmatic interfaces needed for standardized tool wrapping. These challenges are not unique to ToolRosella, but reflect broader difficulties in automated code understanding and environment reproduction~\cite{baker20161}.
Beyond these technical limitations, two further considerations warrant discussion. First, automating the transformation of open-source repositories into callable tools introduces governance and safety challenges. The present work does not include a systematic evaluation of these risks. Section~\ref{sec:security} outlines the relevant design considerations and their implications for future deployment. Second, the present implementation targets Python repositories, leaving substantial scientific code in languages such as R and C++ outside its current coverage. Section~\ref{sec:languages} discusses how the pipeline could be extended across other programming languages.

\subsection{Safety Governance Considerations}
\label{sec:security}

\begin{figure*}[htbp]
  \centering
  \includegraphics[width=\textwidth]{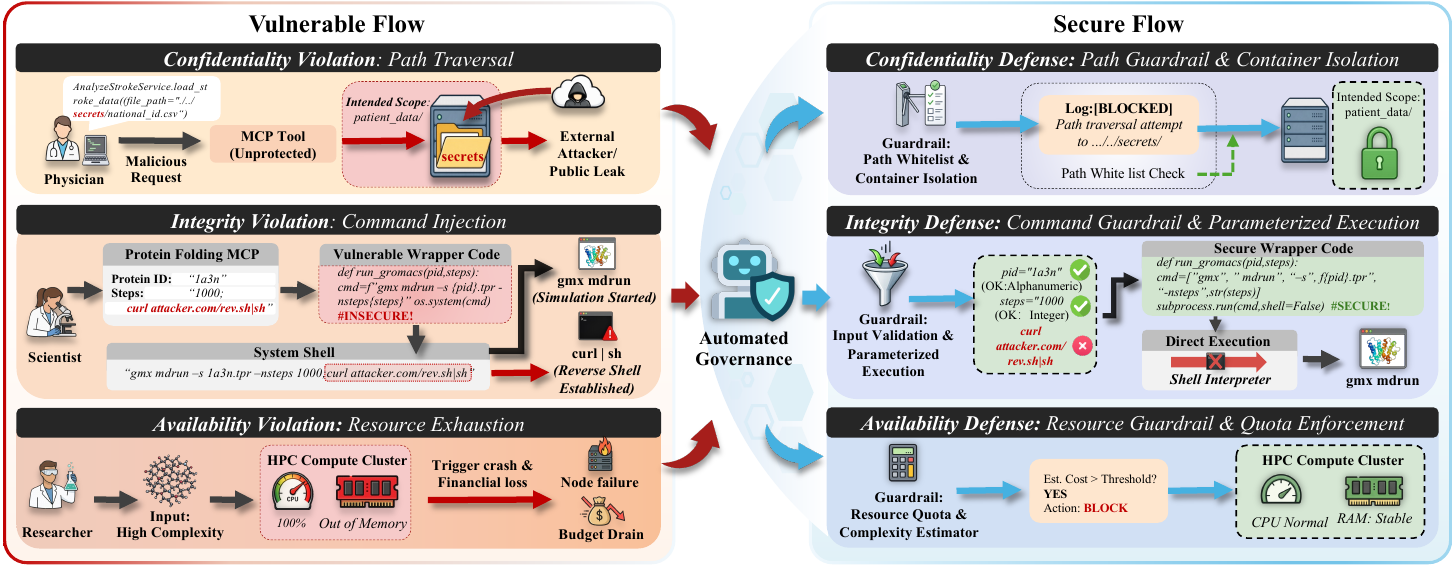}
  \caption{Vulnerable vs. Secure workflow, illustrated with representative cases of (i) data exfiltration in biomedical pipelines, (ii) command injection in legacy tool invocation, and (iii) resource exhaustion in High-Performance Computing (HPC) environments.}
  \label{fig:security_case_cia_flow}
\end{figure*}

Current scientific agent systems often depend on manually curated tool inventories~\cite{castelvecchi2024researchers,chen2026conversational,ha2023ai}. 
ToolRosella expands this capability space by retrieving open-source repositories and transforming their executable capabilities into callable tools, but this expansion shifts the trust boundary from a closed, pre-vetted inventory to externally sourced code that has not been individually reviewed for deployment-time behavior. While the construction pipeline of ToolRosella ensures functional executability, functional executability does not by itself guarantee that invocation will remain within intended operational boundaries. For example, a wrapped tool may still trigger unintended file access, unsafe command execution, or unbounded resource consumption when called by an agent (Fig.~\ref{fig:security_case_cia_flow}). 
We discuss safeguards designed to support at two complementary layers, controls applied when external repositories are wrapped into MCP services, and controls applied when the resulting tools are invoked by agents.


\textbf{Wrapper-level controls.} At the wrapper level, three classes of safeguards target the risks above, corresponding to the confidentiality, integrity, and availability.

\begin{itemize}
    \item Confidentiality: path-bounded execution and isolation. Many scientific workflows operate on sensitive or access-restricted data, including clinical records, biomedical measurements, and genomic sequences~\cite{cordes2024competing,yu2025governance}. When repository-derived functions are exposed as callable tools, file access needs to remain within the intended working scope. At this layer, path whitelisting, restricted working directories, and container isolation provide the confidentiality controls. These mechanisms are intended to constrain file access and reduce the risk of unintended data exposure.
    \item Integrity: constrained invocation and validated interfaces. Scientific repositories often depend on legacy scripts, shell utilities, and heterogeneous command-line programs, creating risk when loosely constrained inputs are passed into executable actions~\cite{kaur2021analysis}. Standardized wrappers help reduce this risk by favoring validated inputs, structured tool interfaces, and controlled invocation of wrapped functionality. Such patterns are designed to preserve interface integrity by ensuring that user-supplied values cannot alter the structure of executed commands.
    \item Availability: resource-aware execution control. Scientific workloads can be highly sensitive to resource consumption, and poorly bounded jobs may exhaust shared compute budgets or destabilize execution environments, especially in HPC settings~\cite{crosby2003denial}. Resource-aware execution policies can combine pre-execution screening with runtime constraints on tool use. Such controls help prevent clearly over-budget or abnormal workloads from being launched.
\end{itemize}

\textbf{Action-level controls.} Beyond wrapper-level safeguards, which govern how each tool is implemented internally, deployment also requires controls over whether and how a given tool is invoked in a particular agent action.
General-purpose, plugin-based agent frameworks such as OpenClaw illustrate an issue. Tools with very different consequences may be presented to the planner through similar invocation interfaces. Standardized ToolRosella wrappers can make tool inputs and invocation formats explicit, creating a control point where additional checks can be applied before higher-impact calls are executed. Low-risk retrieval or analysis tools can be invoked automatically, whereas higher-impact tools can be routed through additional pre-execution checks, such as scope inspection, resource estimation, or user confirmation. In this way, action-level control complements wrapper-level safeguards as a ``safety brake'' during tool use, helping prevent repository-derived tools from being invoked in ways that exceed user intent or deployment boundaries.

\subsection{Extension to other programming languages}
\label{sec:languages}

Scientific computational capabilities are not written in a single language. Although ToolRosella's current implementation targets Python repositories, useful functionality also resides in R, C/C++, Java, and other languages~\cite{baker2017scientific,virtanen2020scipy,harris2020array}. For ToolRosella to serve as a general standardization layer between scientific code and agent execution, its construction pipeline should accommodate these other languages as well. Here, we discuss how the architecture is designed to support such an extension.

The pipeline is structured to separate concerns that are common to all repositories from concerns that are language-specific. Repository retrieval, capability analysis, interface generation, validation, and iterative repair are language-agnostic and apply to any source repository once its callable units are identified. What varies across languages is how these callable units are discovered, built, and invoked. For example, how dependencies are declared, how the build artifacts are produced, and how entry points are exposed. The construction pipeline isolates this variability into language-specific adapters. Adding a new language requires implementing the adapter rather than modifying the pipeline or the tool interface.

Concretely, an adapter for each language would translate that language's native entry points into MCP-compatible tools. For R packages, the adapter would inspect package metadata such as \texttt{DESCRIPTION} and \texttt{NAMESPACE} files, identify exported functions or script entry points, and expose them through structured input-output wrappers. For C and C++ projects, the adapter would use build metadata such as Makefiles or CMake configurations to compile or locate executable targets, then wrap them as Command-Line Interface (CLI) backed MCP tools. For Java projects, the adapter would parse Maven or Gradle build files, locate JAR entry points or service endpoints, and expose selected methods through the same MCP service contract. In each case, the source language determines how functionality is discovered, built, and executed, while the resulting tool interface remains standardized for agent invocation.
This separation is what allows ToolRosella to broaden repository coverage without changing how downstream agents discover, select, and invoke tools.

\section{Methods}
ToolRosella is designed to retrieve task-relevant code repositories, standardize them into callable tools, and orchestrate these tools for user-defined scientific tasks. This process requires coordinated handling of code understanding, environment construction, and interface validation. ToolRosella realizes it through three coordinated components: the \textit{Tool-search Agent}, the \textit{MCP-construction Agent}, and the \textit{Planning Agent}. 
\subsection{Tool-search Agent}

The Tool-search Agent enables ToolRosella to access capabilities beyond a predefined tool pool by retrieving task-relevant repositories from open-source ecosystems.

Given a user query \(Q\), the agent first uses an LLM to infer task intent and extract up to \(m\) topic descriptors, denoted as \(\mathcal{W}=\{W_1,\dots,W_m\}\), that summarize the functional requirements of the request. These descriptors are then used to search for candidate repositories through the GitHub API. The retrieved candidate set is defined as:

\begin{equation}
R_K=\operatorname{TopK}(\operatorname{GitSearch}(\mathcal{W})),
\end{equation}
where \(K\) denotes the number of retrieved repositories. Candidates are ranked primarily by popularity signals (for example, star count), which we use as a practical proxy for community adoption.
The LLM evaluates whether the initial results contain sufficiently relevant candidates. If not, it reformulates the query while preserving its objective. In practice, this reformulation removes overly specific scenario descriptions or constraints, thereby improving retrieval recall without altering task intent.

Retrieved repositories are subsequently evaluated according to two criteria: structural completeness and functional relevance. Structural completeness assesses whether the repository provides sufficient executable assets, including source code, dependency specifications, and setup instructions. Functional relevance measures the extent to which repository capabilities align with the requested task.
The system summarizes these properties in a structured repository assessment and selects the highest-ranked candidate satisfying both criteria. For composite tasks involving multiple subtasks, several complementary repositories can be selected jointly, allowing distinct repositories to contribute specialized functions within a shared workflow.

Unless otherwise specified, we set \(K=50\) and \(m=5\). If a user directly provides a repository name or GitHub link, the retrieval stage is bypassed, and the workflow proceeds to MCP construction.

\subsection{MCP-construction Agent}

The MCP-construction Agent transforms a retrieved GitHub repository into a standardized MCP-compatible service that can be invoked reliably by language models. Because open-source repositories differ substantially in interface design, dependency structure, and execution logic, ToolRosella adopts a staged construction pipeline composed of eight functional nodes: \textit{Download}, \textit{Analysis}, \textit{Env}, \textit{Generate}, \textit{Code\_check}, \textit{Run}, \textit{Review} and \textit{Finish}.

\paragraph{Download Node.}
The selected repository is cloned into a local workspace to enable subsequent parsing, testing, and service construction.

\paragraph{Analysis Node.}
Open-source repositories are commonly organized for human developers rather than autonomous tool invocation, with reusable computational routines often entangled with command-line wrappers, visualization layers, or project-specific execution logic. The Analysis Node addresses this challenge through structured interpretation of repository source code together with accompanying documentation (for example, \textit{README.md}) to decouple reusable functional components, infer callable interfaces, and recover internal dependencies. Because repository-level tool construction requires semantic abstraction beyond raw code parsing, we use \href{https://deepwiki.com}{DeepWiki} in our implementation as a practical utility for codebase-level analysis. This process identifies the roles of individual modules, resolves invocation relationships among components, and aligns documented usage patterns with underlying implementations. Information extracted from source files and documentation is integrated into a structured capability report (Code Report) summarizing the repository’s core functions, interface behavior, and execution logic. The resulting report provides the basis for subsequent environment preparation, service synthesis, and functional validation.

\paragraph{Env Node.}
The Env Node prepares an executable runtime environment for the constructed MCP service by resolving 
dependencies declared in the repository. The node first initializes a Python environment and defaults to \textit{Python 3.10} when no version is specified, ensuring compatibility with MCP service requirements. It then parses dependency specification files (for example, \textit{requirements.txt} or \textit{Dockerfile}) to identify required packages and environment settings. The resolved dependencies are subsequently installed automatically, yielding a reproducible execution environment for downstream validation and deployment.

\paragraph{Generate Node.}
Based on the structured capability report, the Generate Node uses an LLM to synthesize an MCP-compatible service layer that exposes repository functionality through standardized tool interfaces. This stage produces the core service code, interface wrappers, and validation scripts required for deployment. Repository functions are encapsulated within an adapter layer and converted into an asynchronous execution format to support concurrent and non-blocking tool invocation. In addition, the generated templates incorporate predefined rules for parameter conversion, exception handling, and dependency management. These constraints improve interface consistency, type safety, and runtime robustness.

\paragraph{Code\_check Node.}
Because LLM-generated wrappers may contain interface inconsistencies or unsupported references, ToolRosella introduces a Code\_check Node after service synthesis to verify and repair the generated MCP implementation. The primary objective of this stage is to ensure consistency between exposed tool interfaces and the underlying repository code.
This stage first scans source modules \(M\) and extracts publicly accessible functions and classes to construct an available-symbol dictionary \(D\). It then analyzes all import statements in the generated MCP service. A reference is considered valid only if the target module is present in \(D\) and the requested symbol is listed among the exported members of that module.
This validation step helps ensure that imported functions remain executable and that interface bindings are consistent with repository implementations. When inconsistencies are detected, the Code\_check Node initiates an automated repair loop. The system identifies the relevant source files, extracts structural information such as function signatures and class definitions, and combines these signals with execution errors and generated wrapper code. An LLM then revises import statements, interface mappings, and tool definitions to restore consistency between the MCP layer and the underlying repository.
The corrected implementation is subsequently returned for re-validation in later stages.

\paragraph{Run Node.}
The Run Node executes the generated MCP service within the prepared runtime environment and evaluates it using an automated validation suite. This stage verifies whether the synthesized service can be initialized correctly and whether core tool interfaces behave as expected under execution. If all tests pass, the build is marked as successful and proceeds to the Finish Node. If validation fails, execution traces and error logs are captured automatically and forwarded to the Review Node, initiating a corrective cycle of diagnosis and regeneration.

\paragraph{Review Node.}
When validation failures are reported by the Run Node, the Review Node performs targeted diagnosis and repair planning for the generated MCP service. The stage provides the LLM with contextual signals, including execution errors, failing test cases, generated wrapper code, and structural information from the original repository. Based on this evidence, the model identifies likely sources of failure and produces a structured repair plan specifying required code modifications and interface corrections. The repair plan is then returned to the Generate Node to guide targeted regeneration of the MCP service. This iterative feedback mechanism enables progressive correction until the service satisfies downstream validation tests.

\paragraph{Finish Node.}
Once the Run Node confirms that all tests have passed, the system organizes the final version of the MCP service files, the dependency list, usage instructions, and example calls into a standardized directory structure. The resulting service is then consolidated into a deployable MCP package that conforms to community standards.


\subsection{Planning Agent}

After tools are registered, the Planning Agent determines how available services should be invoked to address the user request. This component follows a reasoning-and-action paradigm related to ReAct~\cite{yao2022react}, in which the LLM iteratively interprets the task, selects tools, and updates subsequent actions based on intermediate results.

Given a query \(Q\), the Planning Agent considers task objectives, constraints, and tool descriptions to generate an invocation strategy that maps available tools to subtasks and execution order. If the user issues a refined request \(Q'\), the planner re-evaluates objectives and dynamically updates the workflow.

\section*{Data availability}
The data of standardized repositories and the downstream task evaluation set are available via \url{https://huggingface.co/datasets/ArthurY/ToolRosella}.

\section*{Code availability}
The source code of ToolRosella is available via GitHub at \url{https://github.com/DEFENSE-SEU/ToolRosella}.




\bibliography{sn-bibliography}

@article{bi2023accurate,
 author = {Bi, Kaifeng and Xie, Lingxi and Zhang, Hengheng and Chen, Xin and Gu, Xiaotao and Tian, Qi},
 journal = {Nature},
 number = {7970},
 pages = {533--538},
 publisher = {Nature Publishing Group},
 title = {Accurate medium-range global weather forecasting with 3D neural networks},
 volume = {619},
 year = {2023}
}

@article{ha2023ai,
 author = {Ha, Taesin and Lee, Dongseon and Kwon, Youngchun and Park, Min Sik and Lee, Sangyoon and Jang, Jaejun and Choi, Byungkwon and Jeon, Hyunjeong and Kim, Jeonghun and Choi, Hyundo and others},
 journal = {Science advances},
 number = {44},
 pages = {eadj0461},
 publisher = {American Association for the Advancement of Science},
 title = {AI-driven robotic chemist for autonomous synthesis of organic molecules},
 volume = {9},
 year = {2023}
}

@article{Liu,
 author = {Liu, Han and Li, Liantang},
 langid = {english},
 title = {On Languaging a Simulation Engine}
}

@misc{OpenAgents,
 archiveprefix = {arXiv},
 author = {Tianbao Xie and Fan Zhou and Zhoujun Cheng and Peng Shi and Luoxuan Weng and Yitao Liu and Toh Jing Hua and Junning Zhao and Qian Liu and Che Liu and Leo Z. Liu and Yiheng Xu and Hongjin Su and Dongchan Shin and Caiming Xiong and Tao Yu},
 eprint = {2310.10634},
 primaryclass = {cs.CL},
 title = {OpenAgents: An Open Platform for Language Agents in the Wild},
 year = {2023}
}

@article{shao2025sciscigpt,
 author = {Shao, Erzhuo and Wang, Yifang and Qian, Yifan and Pan, Zhenyu and Liu, Han and Wang, Dashun},
 journal = {arXiv preprint arXiv:2504.05559},
 title = {SciSciGPT: Advancing Human-AI Collaboration in the Science of Science},
 year = {2025}
}

@article{swanson2025virtual,
 author = {Swanson, Kyle and Wu, Wesley and Bulaong, Nicole L and Pak, Jennifer E and Zou, James},
 journal = {Nature},
 pages = {716--723},
 publisher = {Nature Publishing Group},
 title = {The Virtual Lab of AI agents designs new SARS-CoV-2 nanobodies},
 volume = {646},
 year = {2025}
}

@inproceedings{w,
 author = {Ben D. Phillips and Joanna N. Schmidt and Robert Falck and Eliot Aretskin-Hariton},
 booktitle = {AIAA SciTech Forum},
 doi = {10.2514/6.2024-0173},
 month = {January},
 title = {End-To-End Uncertainty Quantification with Analytical Derivatives for Design Under Uncertainty},
 year = {2024}
}

@article{wang2023scientific,
 author = {Wang, Hanchen and Fu, Tianfan and Du, Yuanqi and Gao, Wenhao and Huang, Kexin and Liu, Ziming and Chandak, Payal and Liu, Shengchao and Van Katwyk, Peter and Deac, Andreea and others},
 journal = {Nature},
 number = {7972},
 pages = {47--60},
 publisher = {Nature Publishing Group UK London},
 title = {Scientific discovery in the age of artificial intelligence},
 volume = {620},
 year = {2023}
}

@article{wang2025repomaster,
 author = {Wang, Huacan and others},
 journal = {arXiv preprint arXiv:2505.21577},
 title = {RepoMaster: Autonomous Exploration and Understanding of GitHub Repositories for Complex Task Solving},
 year = {2025}
}

@article{xu2021artificial,
 author = {Xu, Yongjun and Liu, Xin and Cao, Xin and Huang, Changping and Liu, Enke and Qian, Sen and Liu, Xingchen and Wu, Yanjun and Dong, Fengliang and Qiu, Cheng-Wei and others},
 journal = {The Innovation},
 number = {4},
 publisher = {Elsevier},
 title = {Artificial intelligence: A powerful paradigm for scientific research},
 volume = {2},
 year = {2021}
}

@inproceedings{yao2022react,
 author = {Yao, Shunyu and Zhao, Jeffrey and Yu, Dian and Du, Nan and Shafran, Izhak and Narasimhan, Karthik R and Cao, Yuan},
 booktitle = {The eleventh international conference on learning representations},
 title = {React: Synergizing reasoning and acting in language models},
 year = {2022}
}

@article{jumper2021highly,
  title={Highly accurate protein structure prediction with AlphaFold},
  author={Jumper, John and Evans, Richard and Pritzel, Alexander and Green, Tim and Figurnov, Michael and Ronneberger, Olaf and Tunyasuvunakool, Kathryn and Bates, Russ and {\v{Z}}{\'\i}dek, Augustin and Potapenko, Anna and others},
  journal={Nature},
  volume={596},
  number={7873},
  pages={583--589},
  year={2021},
  publisher={Nature Publishing Group UK London}
}

@article{xin2025towards,
  title={Towards agentic science for advancing scientific discovery},
  author={Xin, Hongliang and Kitchin, John R and Kulik, Heather J},
  journal={Nature Machine Intelligence},
  volume={7},
  number={9},
  pages={1373--1375},
  year={2025},
  publisher={Nature Publishing Group UK London}
}

@article{lu2026towards,
  title={Towards end-to-end automation of AI research},
  author={Lu, Chris and Lu, Cong and Lange, Robert Tjarko and Yamada, Yutaro and Hu, Shengran and Foerster, Jakob and Ha, David and Clune, Jeff},
  journal={Nature},
  volume={651},
  number={8107},
  pages={914--919},
  year={2026},
  publisher={Nature Publishing Group UK London}
}

@article{yang2025understanding,
  title={Understanding and manipulating the crystallization of Sn--Pb perovskites for efficient all-perovskite tandem solar cells},
  author={Yang, Xuke and Ma, Tianjun and Hu, Haojun and Ye, Wenjiang and Li, Xin and Li, Mingyu and Zhang, Afei and Ge, Ciyu and Sun, Xianglang and Zhu, Yongxin and others},
  journal={Nature Photonics},
  volume={19},
  number={4},
  pages={426--433},
  year={2025},
  publisher={Nature Publishing Group UK London}
}

@article{Boiko2023,
  title = {Autonomous chemical research with large language models},
  volume = {624},
  ISSN = {1476-4687},
  url = {http://dx.doi.org/10.1038/s41586-023-06792-0},
  DOI = {10.1038/s41586-023-06792-0},
  number = {7992},
  journal = {Nature},
  publisher = {Springer Science and Business Media LLC},
  author = {Boiko,  Daniil A. and MacKnight,  Robert and Kline,  Ben and Gomes,  Gabe},
  year = {2023},
  month = dec,
  pages = {570–578}
}

@article{MBran2024,
  title = {Augmenting large language models with chemistry tools},
  volume = {6},
  ISSN = {2522-5839},
  url = {http://dx.doi.org/10.1038/s42256-024-00832-8},
  DOI = {10.1038/s42256-024-00832-8},
  number = {5},
  journal = {Nature Machine Intelligence},
  publisher = {Springer Science and Business Media LLC},
  author = {M. Bran,  Andres and Cox,  Sam and Schilter,  Oliver and Baldassari,  Carlo and White,  Andrew D. and Schwaller,  Philippe},
  year = {2024},
  month = may,
  pages = {525–535}
}

@article{Ding2025,
  title = {SciToolAgent: a knowledge-graph-driven scientific agent for multitool integration},
  volume = {5},
  ISSN = {2662-8457},
  url = {http://dx.doi.org/10.1038/s43588-025-00849-y},
  DOI = {10.1038/s43588-025-00849-y},
  number = {10},
  journal = {Nature Computational Science},
  publisher = {Springer Science and Business Media LLC},
  author = {Ding,  Keyan and Yu,  Jing and Huang,  Junjie and Yang,  Yuchen and Zhang,  Qiang and Chen,  Huajun},
  year = {2025},
  month = aug,
  pages = {962–972}
}

@inproceedings{lyu-etal-2025-enhancing,
    title = "Enhancing Open-Domain Task-Solving Capability of {LLM}s via Autonomous Tool Integration from {G}it{H}ub",
    author = "Lyu, Bohan  and
      Cong, Xin  and
      Yu, Heyang  and
      Yang, Pan  and
      Qian, Cheng  and
      Wang, Zihe  and
      Qin, Yujia  and
      Ye, Yining  and
      Lu, Yaxi  and
      Qian, Chen  and
      Zhang, Zhong  and
      Yan, Yukun  and
      Lin, Yankai  and
      Liu, Zhiyuan  and
      Sun, Maosong",
    editor = "Che, Wanxiang  and
      Nabende, Joyce  and
      Shutova, Ekaterina  and
      Pilehvar, Mohammad Taher",
    booktitle = "Proceedings of the 63rd Annual Meeting of the Association for Computational Linguistics (Volume 1: Long Papers)",
    month = jul,
    year = "2025",
    address = "Vienna, Austria",
    publisher = "Association for Computational Linguistics",
    url = "https://aclanthology.org/2025.acl-long.845/",
    doi = "10.18653/v1/2025.acl-long.845",
    pages = "17257--17277",
    ISBN = "979-8-89176-251-0"
}

@article{kelly2024age,
  title={Age-and sex-specific analysis of stroke hospitalization rates, risk factors, and outcomes from German nationwide data},
  author={Kelly, Dearbhla M and Engelbertz, Christiane and Rothwell, Peter M and Anderson, Christopher D and Reinecke, Holger and Koeppe, Jeanette},
  journal={Stroke},
  volume={55},
  number={9},
  pages={2284--2294},
  year={2024},
  publisher={Lippincott Williams \& Wilkins Hagerstown, MD}
}

@article{howard2025association,
  title={Association of duration of recognized hypertension and stroke risk: The REGARDS study},
  author={Howard, George and Muntner, Paul and Lackland, Daniel T and Plante, Timothy B and Cushman, Mary and Stamm, Brian and Judd, Suzanne E and Howard, Virginia J},
  journal={Stroke},
  volume={56},
  number={1},
  pages={105--112},
  year={2025},
  publisher={Lippincott Williams \& Wilkins Hagerstown, MD}
}

@article{xu2022fusarium,
  title={Fusarium fruiting body microbiome member Pantoea agglomerans inhibits fungal pathogenesis by targeting lipid rafts},
  author={Xu, Sunde and Liu, Yong-Xin and Cernava, Tomislav and Wang, Hongkai and Zhou, Yaqi and Xia, Tie and Cao, Shugeng and Berg, Gabriele and Shen, Xing-Xing and Wen, Ziyue and others},
  journal={Nature Microbiology},
  volume={7},
  number={6},
  pages={831--843},
  year={2022},
  publisher={Nature Publishing Group UK London}
}

@article{klughammer2023comparative,
  title={Comparative analysis of genome-scale, base-resolution DNA methylation profiles across 580 animal species},
  author={Klughammer, Johanna and Romanovskaia, Daria and Nemc, Amelie and Posautz, Annika and Seid, Charlotte A and Schuster, Linda C and Keinath, Melissa C and Lugo Ramos, Juan Sebastian and Kosack, Lindsay and Evankow, Ann and others},
  journal={Nature Communications},
  volume={14},
  number={1},
  pages={232},
  year={2023},
  publisher={Nature Publishing Group UK London}
}

@article{li2025stabilizing,
  title={Stabilizing high-efficiency perovskite solar cells via strategic interfacial contact engineering},
  author={Li, Guixiang and Zhang, Zuhong and Agyei-Tuffour, Benjamin and Wu, Luyan and Gries, Thomas W and Prashanthan, Karunanantharajah and Musiienko, Artem and Li, Jinzhao and Zhu, Rui and Hart, Lucy JF and others},
  journal={Nature Photonics},
  pages={1--8},
  year={2025},
  publisher={Nature Publishing Group UK London}
}

@article{wu2025resilience,
  title={Resilience pathways for halide perovskite photovoltaics under temperature cycling},
  author={Wu, Luyan and Hu, Shuaifeng and Yang, Feng and Li, Guixiang and Wang, Junke and Zuo, Weiwei and Jer{\'o}nimo-Rendon, Jos{\'e} J and Turren-Cruz, Silver-Hamill and Saba, Michele and Saliba, Michael and others},
  journal={Nature Reviews Materials},
  volume={10},
  number={7},
  pages={536--549},
  year={2025},
  publisher={Nature Publishing Group UK London}
}

@article{hayes2025simulating,
  title={Simulating 500 million years of evolution with a language model},
  author={Hayes, Thomas and Rao, Roshan and Akin, Halil and Sofroniew, Nicholas J and Oktay, Deniz and Lin, Zeming and Verkuil, Robert and Tran, Vincent Q and Deaton, Jonathan and Wiggert, Marius and others},
  journal={Science},
  volume={387},
  number={6736},
  pages={850--858},
  year={2025},
  publisher={American Association for the Advancement of Science}
}

@article{hatfield2021data,
  title={The data-driven future of high-energy-density physics},
  author={Hatfield, Peter W and Gaffney, Jim A and Anderson, Gemma J and Ali, Suzanne and Antonelli, Luca and Ba{\c{s}}e{\u{g}}mez du Pree, Suzan and Citrin, Jonathan and Fajardo, Marta and Knapp, Patrick and Kettle, Brendan and others},
  journal={Nature},
  volume={593},
  number={7859},
  pages={351--361},
  year={2021},
  publisher={Nature Publishing Group UK London}
}

@article{qu2024promoting,
  title={Promoting interactions between cognitive science and large language models},
  author={Qu, Youzhi and Du, Penghui and Che, Wenxin and Wei, Chen and Zhang, Chi and Ouyang, Wanli and Bian, Yatao and Xu, Feiyang and Hu, Bin and Du, Kai and others},
  journal={The Innovation},
  volume={5},
  number={2},
  year={2024},
  publisher={Elsevier}
}

@article{woelfle2011open,
  title={Open science is a research accelerator},
  author={Woelfle, Michael and Olliaro, Piero and Todd, Matthew H},
  journal={Nature chemistry},
  volume={3},
  number={10},
  pages={745--748},
  year={2011},
  publisher={Nature Publishing Group UK London}
}

@article{di2025stop,
  title={Stop treating code like an afterthought: record, share and value it},
  author={Di Cosmo, Roberto and Granger, Sabrina and Hinsen, Konrad and Jullien, Nicolas and Le Berre, Daniel and Louvet, Violaine and Maumet, Camille and Maurice, Cl{\'e}mentine and Monat, Rapha{\"e}l and Rougier, Nicolas P},
  journal={Nature},
  volume={646},
  number={8084},
  pages={284--286},
  year={2025},
  publisher={Nature Publishing Group UK London}
}

@article{van2024encore,
  title={ENCORE: a practical implementation to improve reproducibility and transparency of computational research},
  author={van Kampen, Antoine HC and Mahamune, Utkarsh and Jongejan, Aldo and van Schaik, Barbera DC and Balashova, Daria and Lashgari, Danial and Pras-Raves, Mia and Wever, Eric JM and Dane, Adrie D and Garc{\'\i}a-Valiente, Rodrigo and others},
  journal={Nature Communications},
  volume={15},
  number={1},
  pages={8117},
  year={2024},
  publisher={Nature Publishing Group UK London}
}

@article{yang2024swe,
  title={Swe-agent: Agent-computer interfaces enable automated software engineering},
  author={Yang, John and Jimenez, Carlos E and Wettig, Alexander and Lieret, Kilian and Yao, Shunyu and Narasimhan, Karthik and Press, Ofir},
  journal={Advances in Neural Information Processing Systems},
  volume={37},
  pages={50528--50652},
  year={2024}
}

@article{tee2022framework,
  title={A framework for tool cognition in robots without prior tool learning or observation},
  author={Tee, Keng Peng and Cheong, Samuel and Li, Jun and Ganesh, Gowrishankar},
  journal={Nature Machine Intelligence},
  volume={4},
  number={6},
  pages={533--543},
  year={2022},
  publisher={Nature Publishing Group UK London}
}

@article{vriza2026operating,
  title = {Operating advanced scientific instruments with AI agents that learn on the job},
  volume = {12},
  ISSN = {2057-3960},
  url = {http://dx.doi.org/10.1038/s41524-026-02005-0},
  DOI = {10.1038/s41524-026-02005-0},
  number = {1},
  journal = {npj Computational Materials},
  publisher = {Springer Science and Business Media LLC},
  author = {Vriza,  Aikaterini and Prince,  Michael H. and Zhou,  Tao and Chan,  Henry and Cherukara,  Mathew J.},
  year = {2026},
  month = Mar 
}

@article{lin2023evolutionary,
  title={Evolutionary-scale prediction of atomic-level protein structure with a language model},
  author={Lin, Zeming and Akin, Halil and Rao, Roshan and Hie, Brian and Zhu, Zhongkai and Lu, Wenting and Smetanin, Nikita and Verkuil, Robert and Kabeli, Ori and Shmueli, Yaniv and others},
  journal={Science},
  volume={379},
  number={6637},
  pages={1123--1130},
  year={2023},
  publisher={American Association for the Advancement of Science}
}

@article{chen2026conversational,
  title = {A conversational multi-agent AI system for automated plant phenotyping},
  ISSN = {2041-1723},
  url = {http://dx.doi.org/10.1038/s41467-026-71090-y},
  DOI = {10.1038/s41467-026-71090-y},
  journal = {Nature Communications},
  publisher = {Springer Science and Business Media LLC},
  author = {Chen,  Feng and Stogiannidis,  Ilias and Wood,  Andrew and Bueno,  Danilo and Williams,  Dominic and Macfarlane,  Fraser and Grieve,  Bruce D. and Wells,  Darren and Atkinson,  Jonathan A. and Hawkesford,  Malcolm J. and Rolfe,  Stephen A. and Lawson,  Tracy and Pridmore,  Tony and Tsaftaris,  Sotirios A. and Giuffrida,  Mario Valerio},
  year = {2026},
  month = Apr 
}

@article{castelvecchi2024researchers,
  title={Researchers built an ‘AI Scientist’—what can it do},
  author={Castelvecchi, Davide},
  journal={Nature},
  volume={633},
  number={8029},
  pages={266--266},
  year={2024},
  publisher={Nature}
}

@article{qu2026crispr,
  title={CRISPR-GPT for agentic automation of gene-editing experiments},
  author={Qu, Yuanhao and Huang, Kaixuan and Yin, Ming and Zhan, Kanghong and Liu, Dyllan and Yin, Di and Cousins, Henry C and Johnson, William A and Wang, Xiaotong and Shah, Mihir and others},
  journal={Nature Biomedical Engineering},
  volume={10},
  number={2},
  pages={245--258},
  year={2026},
  publisher={Nature Publishing Group UK London}
}

@article{yang2026tool,
  title={Tool-wielding language model-based agent offers conversational exploration of clinical tabular data},
  author={Yang, Andrew and Woo, Joshua and Zhang, Ryan and Mach, Alan and Ramkumar, Prem and Ma, Ying},
  journal={npj Artificial Intelligence},
  volume={2},
  number={1},
  pages={22},
  year={2026},
  publisher={Nature Publishing Group UK London}
}

@article{sacco2024prevention,
  title={Prevention and treatment of ischaemic and haemorrhagic stroke in people with diabetes mellitus: a focus on glucose control and comorbidities},
  author={Sacco, Simona and Foschi, Matteo and Ornello, Raffaele and De Santis, Federico and Pofi, Riccardo and Romoli, Michele},
  journal={Diabetologia},
  volume={67},
  number={7},
  pages={1192--1205},
  year={2024},
  publisher={Springer}
}

@article{zhu2025global,
  title={Global and regional burden of ischemic stroke disease from 1990 to 2021: an age-period-cohort analysis},
  author={Zhu, Weimin and He, Xiaxia and Huang, Daochao and Jiang, Yiqing and Hong, Weijun and Ke, Shaofa and Wang, En and Wang, Feng and Wang, Xianwei and Shan, Renfei and others},
  journal={Translational Stroke Research},
  volume={16},
  number={5},
  pages={1474--1485},
  year={2025},
  publisher={Springer}
}

@misc{baker20161,
  title = {1, 500 scientists lift the lid on reproducibility},
  volume = {533},
  ISSN = {1476-4687},
  DOI = {10.1038/533452a},
  number = {7604},
  journal = {Nature},
  publisher = {Springer Science and Business Media LLC},
  author = {Baker,  Monya},
  year = {2016},
  month = May,
  pages = {452–454}
}

@article{virtanen2020scipy,
  title={SciPy 1.0: fundamental algorithms for scientific computing in Python},
  author={Virtanen, Pauli and Gommers, Ralf and Oliphant, Travis E and Haberland, Matt and Reddy, Tyler and Cournapeau, David and Burovski, Evgeni and Peterson, Pearu and Weckesser, Warren and Bright, Jonathan and others},
  journal={Nature methods},
  volume={17},
  number={3},
  pages={261--272},
  year={2020},
  publisher={Nature Publishing Group US New York}
}

@article{harris2020array,
  title={Array programming with NumPy},
  author={Harris, Charles R and Millman, K Jarrod and Van Der Walt, St{\'e}fan J and Gommers, Ralf and Virtanen, Pauli and Cournapeau, David and Wieser, Eric and Taylor, Julian and Berg, Sebastian and Smith, Nathaniel J and others},
  journal={nature},
  volume={585},
  number={7825},
  pages={357--362},
  year={2020},
  publisher={Nature Publishing Group UK London}
}

@article{baker2017scientific,
  title={Scientific computing: Code alert},
  author={Baker, Monya},
  journal={Nature},
  volume={541},
  number={7638},
  pages={563--565},
  year={2017},
  publisher={Nature Publishing Group UK London}
}

@article{wang2026ai,
  title = {AI-driven complex systems redefine cognitive science},
  volume = {7},
  ISSN = {2666-6758},
  url = {http://dx.doi.org/10.1016/j.xinn.2026.101314},
  DOI = {10.1016/j.xinn.2026.101314},
  number = {5},
  journal = {The Innovation},
  publisher = {Elsevier BV},
  author = {Wang,  Peng and Sun,  Xu and Zou,  Liye and Law,  Effie Lai-Chong and Paas,  Fred},
  year = {2026},
  month = May,
  pages = {101314}
}

@inproceedings{crosby2003denial,
  author = {Scott A. Crosby and Dan S. Wallach},
  title = {Denial of Service via Algorithmic Complexity Attacks},
  booktitle = {12th USENIX Security Symposium (USENIX Security 03)},
  year = {2003},
  address = {Washington, D.C.},
  url = {https://www.usenix.org/conference/12th-usenix-security-symposium/denial-service-algorithmic-complexity-attacks},
  publisher = {USENIX Association},
  month = aug
}

@article{kaur2021analysis,
  title={An analysis of security vulnerabilities in container images for scientific data analysis},
  author={Kaur, Bhupinder and Dugr{\'e}, Mathieu and Hanna, Aiman and Glatard, Tristan},
  journal={GigaScience},
  volume={10},
  number={6},
  pages={giab025},
  year={2021},
  publisher={Oxford University Press}
}

@article{yu2025governance,
  title={Governance of cross-border genomic data sharing through a human rights approach},
  author={Yu, Liang and Feng, Ruohan and Sun, Youhai and Peng, Yaojin},
  journal={Nature Genetics},
  volume={57},
  number={9},
  pages={2090--2098},
  year={2025},
  publisher={Nature Publishing Group US New York}
}

@article{cordes2024competing,
  title={Competing interests: digital health and indigenous data sovereignty},
  author={Cordes, Ashley and Bak, Marieke and Lyndon, Mataroria and Hudson, Maui and Fiske, Amelia and Celi, Leo Anthony and McLennan, Stuart},
  journal={NPJ digital medicine},
  volume={7},
  number={1},
  pages={178},
  year={2024},
  publisher={Nature Publishing Group UK London}
}

@article{ferber2025development,
  title={Development and validation of an autonomous artificial intelligence agent for clinical decision-making in oncology},
  author={Ferber, Dyke and El Nahhas, Omar SM and W{\"o}lflein, Georg and Wiest, Isabella C and Clusmann, Jan and Le{\ss}mann, Marie-Elisabeth and Foersch, Sebastian and Lammert, Jacqueline and Tschochohei, Maximilian and J{\"a}ger, Dirk and others},
  journal={Nature cancer},
  volume={6},
  number={8},
  pages={1337--1349},
  year={2025},
  publisher={Nature Publishing Group US New York}
}

\section*{Acknowledgements}
This work is funded by Basic Research Program of Jiangsu under Grant No. BK20253021 and National Science Foundation of China (NSFC) under Grant No. 62506075 and 62225602.
We also thank the open-source communities behind the GitHub repositories used in this study, whose work made ToolRosella possible.

\section*{Author contributions}
S.D. conceived the study. X.Y. and H.G. implemented the overall framework and conducted the experiments. C.O. contributed to the safety considerations and multi-language extension. Y.L. and L.Y. revised the manuscript. Z.C. produced the figures. J.Z. and S.P. supported the case studies and provided domain expertise. L.Z., J.Y., Y.R., and M.-L.Z. supervised the entire project. 
All authors wrote the paper, reviewed it, and approved the final paper.

\section*{Competing interests}
The authors declare no competing interests.

\end{document}